\documentclass[aps,prc,superscriptaddress,showpacs,amssymb,amsmath,amsfonts,twocolumn,floatfix,nofootinbib]{revtex4-1}
\usepackage{graphicx}

\newcommand{\be}{\begin{equation}}
\newcommand{\ee}{\end{equation}}
\newcommand{\ba}{\begin{array}}
\newcommand{\ea}{\end{array}}

\newcommand{\GWU}{Data Analysis Center at the Institute for Nuclear 
	Studies, Department of Physics,
        The George Washington University, Washington, D.C. 20052}
\begin{document}

\title{Parameterization dependence of $T$ matrix poles and eigenphases \\ 
       from a fit to $\pi N$ elastic scattering data}

\author{R.~L.~Workman}
\affiliation{\GWU}
\author{R.~A.~Arndt$^\ast$}
\affiliation{\GWU}
\author{W.~J.~Briscoe}
\affiliation{\GWU}
\author{M.~W.~Paris$^\dagger$}
\affiliation{\GWU}
\author{I.~I.~Strakovsky}
\affiliation{\GWU}

\vspace{5cm}
\date{\today}

\begin{abstract}
We compare fits to $\pi N$ elastic scattering
data, based on a Chew-Mandelstam $K$-matrix formalism. Resonances, 
characterized by $T$-matrix poles, are compared in fits generated with 
and without explicit Chew-Mandelstam $K$-matrix poles. Diagonalization 
of the $S$ matrix yields the eigenphase representation. While the 
eigenphases can vary significantly for the different parameterizations, 
the locations of most $T$-matrix poles are relatively stable.   
\end{abstract}

\maketitle

\section{Introduction}
\label{sec:intro}

The excited states of the nucleon~\cite{PDG} have been studied in a wide
array of reactions initiated mainly by pion and photon beams.
Other approaches have involved an examination of the invariant
mass distribution of products from, for example, nucleon-nucleon
reactions~\cite{morsch} and $J/ \Psi$ decays~\cite{bes}. 
Most states listed by the
PDG~\cite{PDG} were identified from fits to $\pi N$ elastic
scattering and reaction data. Photo-decay amplitudes were
determined mostly through analyses of single-pion photoproduction
data. 

Recent measurements of cross section and polarization quantities,
related to the photo- and electroproduction of states other than
$\pi N$, have been analyzed separately and in multi-channel
approaches.  These studies have provided stronger evidence for
states seen only weakly in $\pi N$ elastic scattering, and have
suggested new states, coupling more strongly to other 
channels~\cite{review}.

Among the most extensive $\pi N$ scattering analyses
\cite{KH80,CMB,Arndt2006}, the parametrization of Ref.\cite{Arndt2006}
based on the SAID interactive fitting and database
codes\cite{SAID-website} (the SAID-GW fit), utilizing the most
recent data, has found the fewest number of $N$ and $\Delta$
resonances. In the fit of Ref.~\cite{sm95}, a search for weaker
structures was carried out. There, the existing solution was
modified using a simple product S-matrix approach, to include the
effect of an added Breit-Wigner resonance in each partial wave.
Chi-squared was mapped for various combination of masses, widths
and branching fraction. Two marginally significant candidates
were found in the $S_{11}$ and $F_{15}$ partial waves, with pole
positions: $1689-i96$~MeV (for $S_{11}$) and $1793-i94$ (for
$F_{15}$). Of these, the $F_{15}$ has been reported in subsequent
fits, while the $S_{11}$ has not.

\begin{minipage}{8cm}
\vspace{2.5cm}
$^\ast$Deceased\\
$^\dagger$Current address: Theory Division, 
Los Alamos National Laboratory, Los 
	Alamos, NM 87545, USA
\end{minipage}

Here we have considered another approach. As detailed further
below, existing GW-SAID fits to $\pi N$ elastic scattering data
have utilized a fit form based on the Chew-Mandelstam (CM)
$K$-matrix. This approach is capable of generating $T$-matrix poles
without the assumption of explicit $K$-matrix poles. Previous fits
have only included an explicit CM $K$-matrix pole for the $\Delta
(1232)$. In the present study, an alternative parametrization
with one explicit CM $K$-matrix pole in each partial wave is used
to generate a fit independent of the usual CM parametrization
form.

The motivation for this new fit is twofold. By changing the
parameterization, we are able to gauge the stability of the 
amplitudes and resonance positions. We are also able to see if 
the addition of new explicit CM $K$-matrix poles translates into 
additional resonance signals. 

Below, in Sec.~\ref{sec:form}, we briefly review the CM $K$-matrix 
formalism used in this and previous fits. The eigenphase 
representation, and some numerical details, are reviewed in 
Sec.~\ref{sec:egnph}. Results for the partial wave fits and 
resonance spectrum are compared in Sec.~\ref{sec:results}. 
Finally, in Sec.~\ref{sec:concl}, we consider the implication of 
this and future work.

\section{Chew-Mandelstam Formalism}
\label{sec:form}

The Chew-Mandelstam approach for the parametrization of
multichannel hadronic $\pi N$ elastic scattering and reactions to
other hadronic channels has been described in detail in
Refs.~\cite{sm95,Arndt85,Arndt03,Arndt2006,
Paris:2010tz}. The $\chi^2$-fits to data have been additionally
constrained using the forward $C^{\pm}$ dispersion relations and
fixed-t dispersion relations for the invariant $B$ amplitudes. 

The standard CM parametrization can be expressed in terms of the
on-shell Heitler partial wave $K$ matrix, $K$ as
\begin{equation}
\label{eqn:HKCMK}
	K^{-1}(E) = \overline{K}^{\; -1}(E) - {\rm Re} C(E),
\end{equation}
where $E$ is the (complex) scattering energy, $\overline{K}$ is the
Chew-Mandelstam $K$ matrix and $C$ is a diagonal matrix, whose matrix
elements are termed the Chew-Mandelstam
functions~\cite{Basdevant}. The Heitler $K$ matrix is related to the
partial wave transition amplitude matrix, $T$ as
\begin{equation}
	T^{-1}(E) = K^{-1}(E) - i\rho(E).
\end{equation}
Here, $\rho(E) = \delta(E-H_0)$, where $H_0$ is the (model
independent) relativistic free-particle Hamiltonian with physical
(stable) particle masses. It determines the CM functions, $C(E)$ 
via the relation ${\rm Im} C(E) =\rho(E)$.\footnote{The
inclusion of quasi-two-body channels, such as $\pi\Delta$ are
constrained to have branch points at the stable three-body
thresholds.}

The standard form used in the GW fits is defined by the choice for 
the CM $K$-matrix elements
\begin{equation}
\label{eqn:Kmatpar}
	\overline{K}(E) = \sum_n c_n \overline{z}^n(E),
\end{equation}
where $c_n$ are a set of constants and $\overline{z}$ is a linear 
function of the scattering energy, $E$.  The integer, $n$ is 
typically between 2 and 5, and depends on the matrix element in 
question.

Note that $\overline{K}$ defines an entire function of the complex
parameter $E$ for finite values. This form is used for all but
the $P_{33}$ partial wave, which includes an explicit pole in
$\overline{K}$. For partial waves other than the $P_{33}$, we see that
the CM $K$ matrix, $\overline{K}$, is without poles (or other
singularities). The Heitler $K$ matrix, $K$, 
\begin{equation}
	K = {1\over {1-\overline{K} [{\rm Re C}]}}\overline{K}
\end{equation} 
has a pole whenever $\det[1- {\rm Re} C(E)\overline{K}(E)]=0$. The
matrices $K$ and $\overline{K}$ are free of branch point
singularities~\cite{Zimmerman:1961aa,Paris:2010tz}.

The alternate form of the CM $K$ matrix is similar to the form used 
in the $P_{33}$ partial wave of the standard $\overline{K}$ 
parametrization, described above. This form is given by
\begin{equation}
\label{eqn:Kbars}
	\overline{K}_{ij} = \frac{\gamma_i \gamma_j}{E -E_p} 
	+ \beta(E)_{ij}.
\end{equation}
Here, $\gamma_i(E)$ is a polynomial without a zero at the pole 
position, $E_p$, and the index labels the channel ($\pi N$, $\pi 
\Delta$, $\rho N$, and $\eta N$), $\beta(E)$ is an entire function 
of the complex energy, $E$. 

\section{Eigenphase representation}
\label{sec:egnph}

The fit produces a unitary S-matrix of amplitudes for all 
contributing channels. While those channels not fitted to data 
are unlikely to give a quantitative representation of the reaction 
(for example $\pi N \to\pi\Delta$), they can be used to construct 
a set of eigenphases, which provide an interesting characterization 
of resonance behavior. 

The unitarity of the $S$ matrix implies that its eigenvalues are
phase factors.  The matrix, $U$, of eigenvectors diagonalizes the 
$S$ matrix as
\begin{align}
\label{eqn:diagS}
	U^\dag S U &= \lambda,
\end{align}
where
\begin{align}
	\lambda &=
	\left(\begin{matrix}
	\lambda_1 & 0           & \cdots & 0 \\
	0         & \lambda_2   & \cdots & 0 \\
	0         & 0           & \ddots & 0 \\
	0         & 0           & \cdots & \lambda_n
	\end{matrix}\right) .
\end{align}

Exploiting $| \lambda_i |=1$ we write
\begin{align}
	\lambda_i &= e^{2i\phi_i}
\end{align}
with $\phi_i$ real. 

Our objective is the numerical evaluation of the eigenphases
given the $T$-matrix elements from various fits.  This is
straightforward at a given energy, using a standard routine to
diagonalize the unitary $S$ matrix.  The only complicating issue
is correlating a given eigenphase, $\phi_i(E)$ with the
appropriate eigenchannel when two (or more) eigenphases converge 
as the energy changes. In other words, once an eigenchannel $i$ 
is determined, we must track it for all energies. The no-crossing
theorem\cite{Wigner:1929nc} is readily generalized to unitary
matrices and shows that, in a given partial wave, the eigenphases
may not be equal for any energy. This property is exhibited
in the eigenphase plots discussed below.

\begin{figure*}[ht]
\centerline{
\includegraphics[height=0.42\textwidth, angle=90]{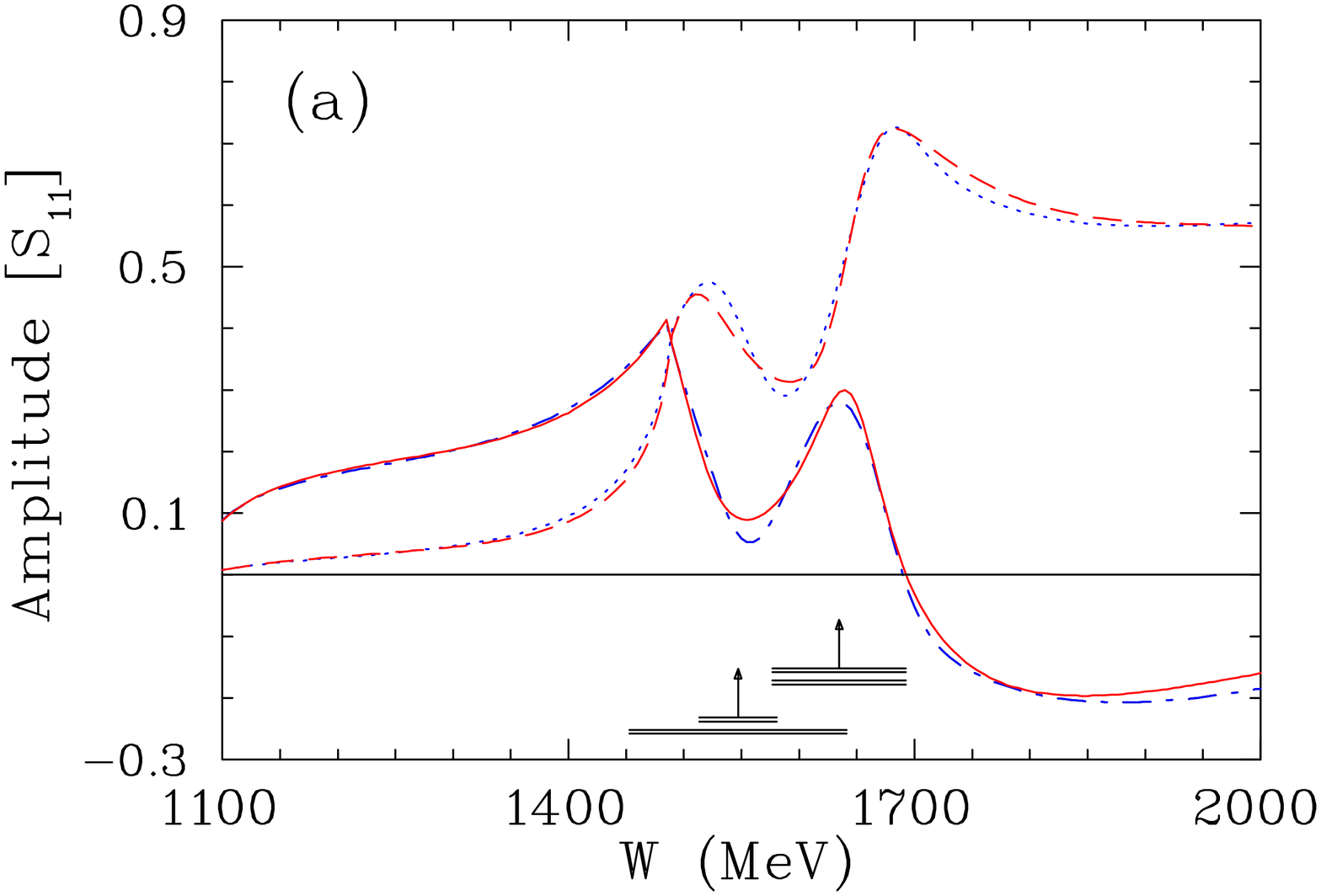}\hfill
\includegraphics[height=0.42\textwidth, angle=90]{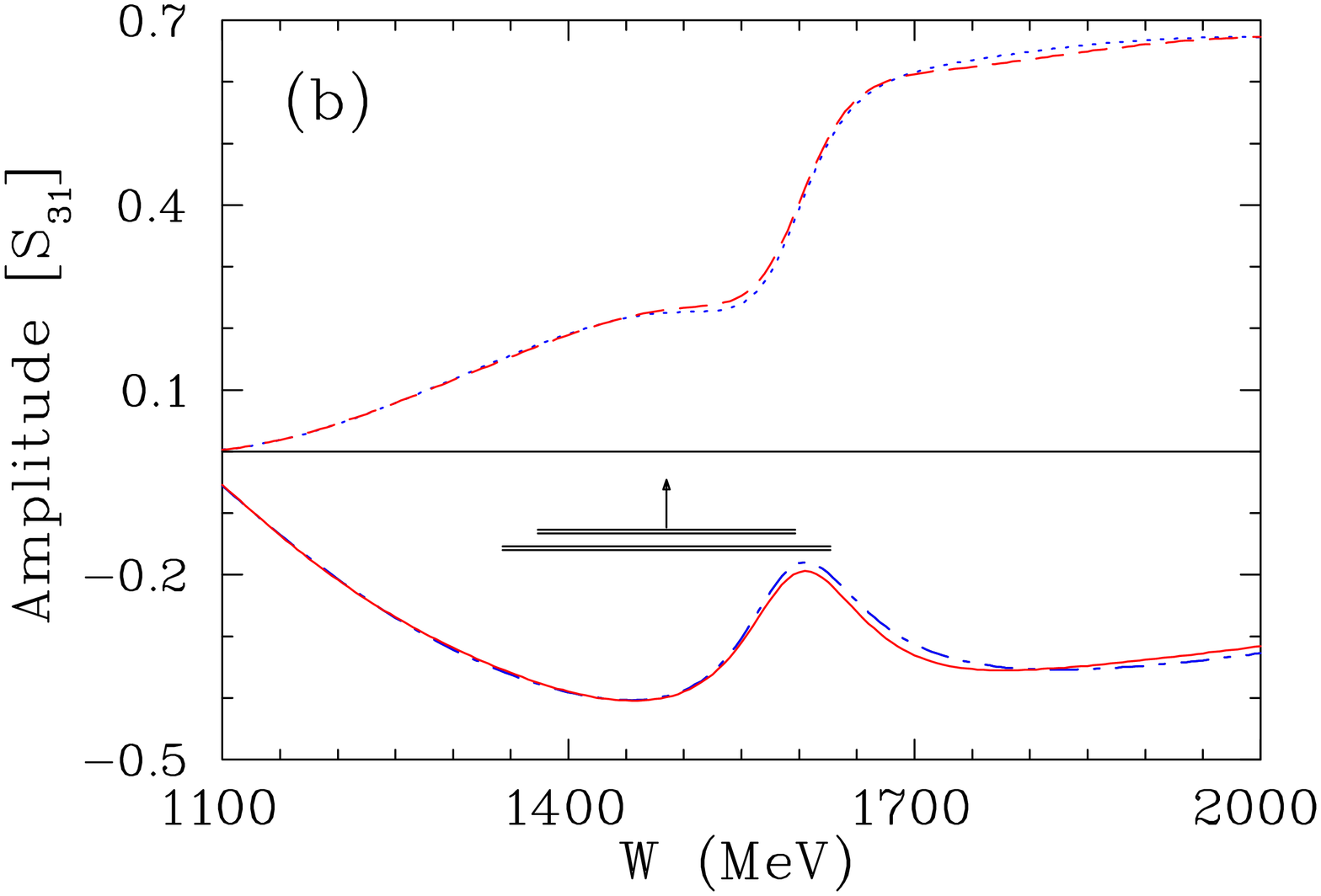}}
\centerline{
\includegraphics[height=0.42\textwidth, angle=90]{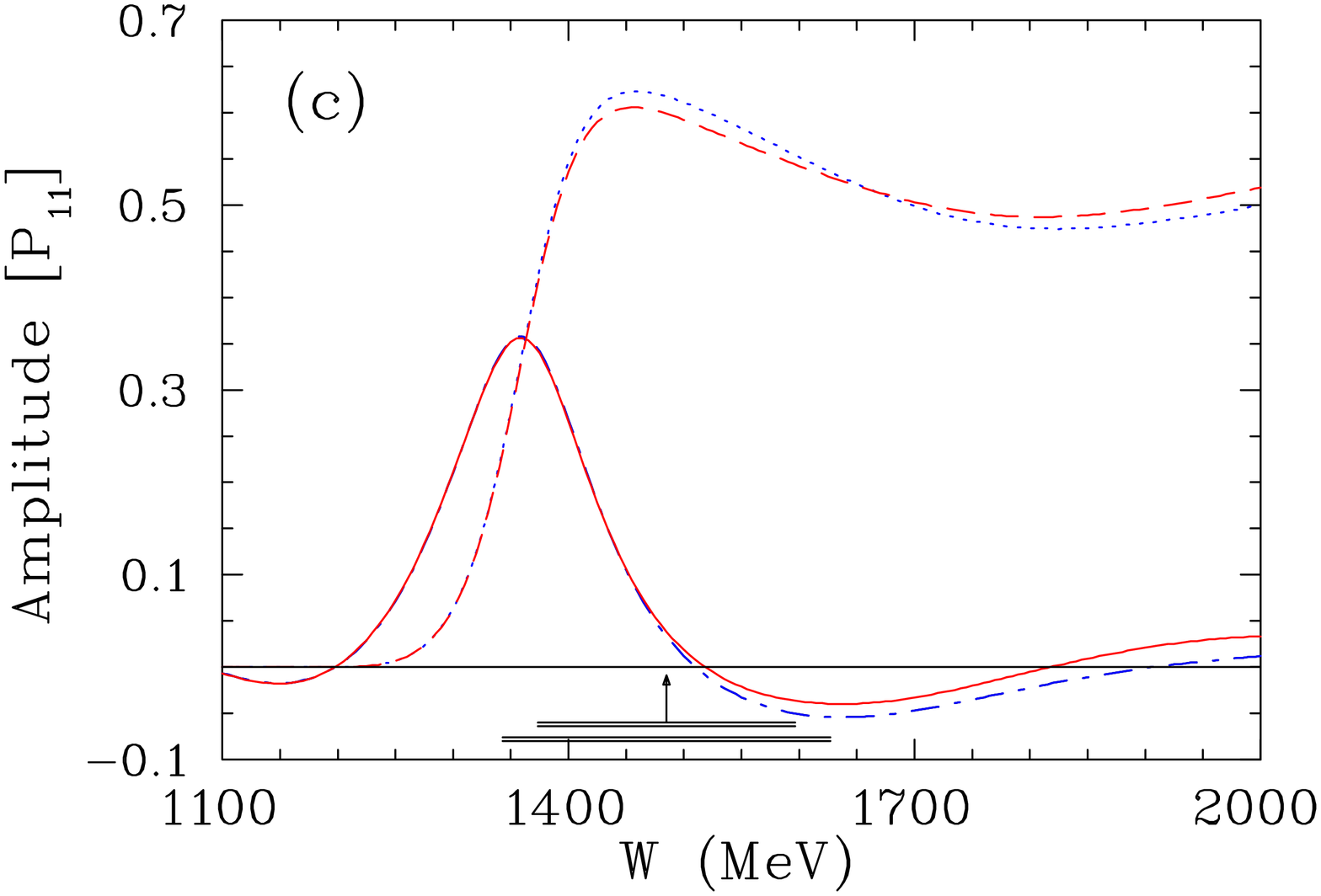}\hfill
\includegraphics[height=0.42\textwidth, angle=90]{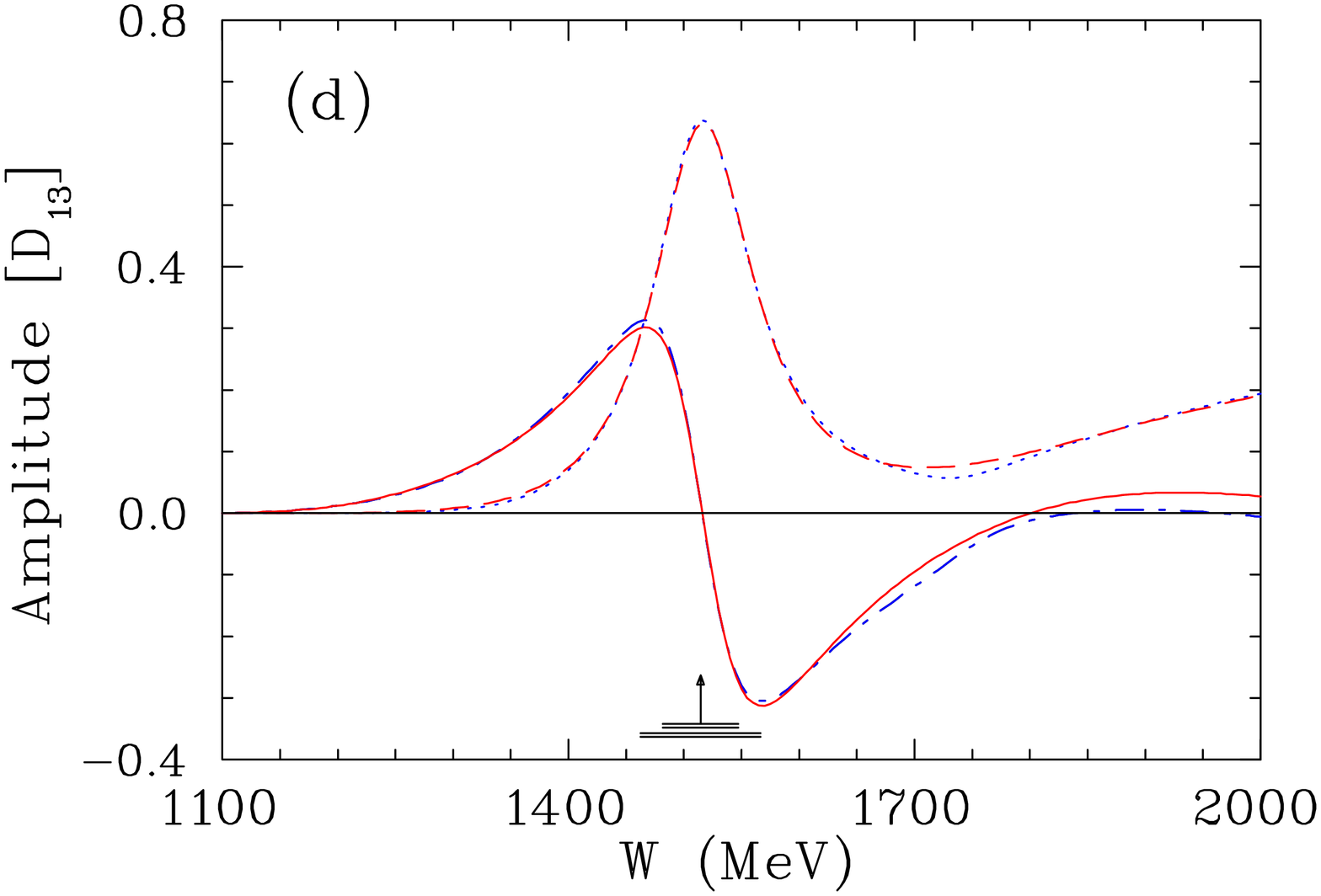}}
\centerline{
\includegraphics[height=0.42\textwidth, angle=90]{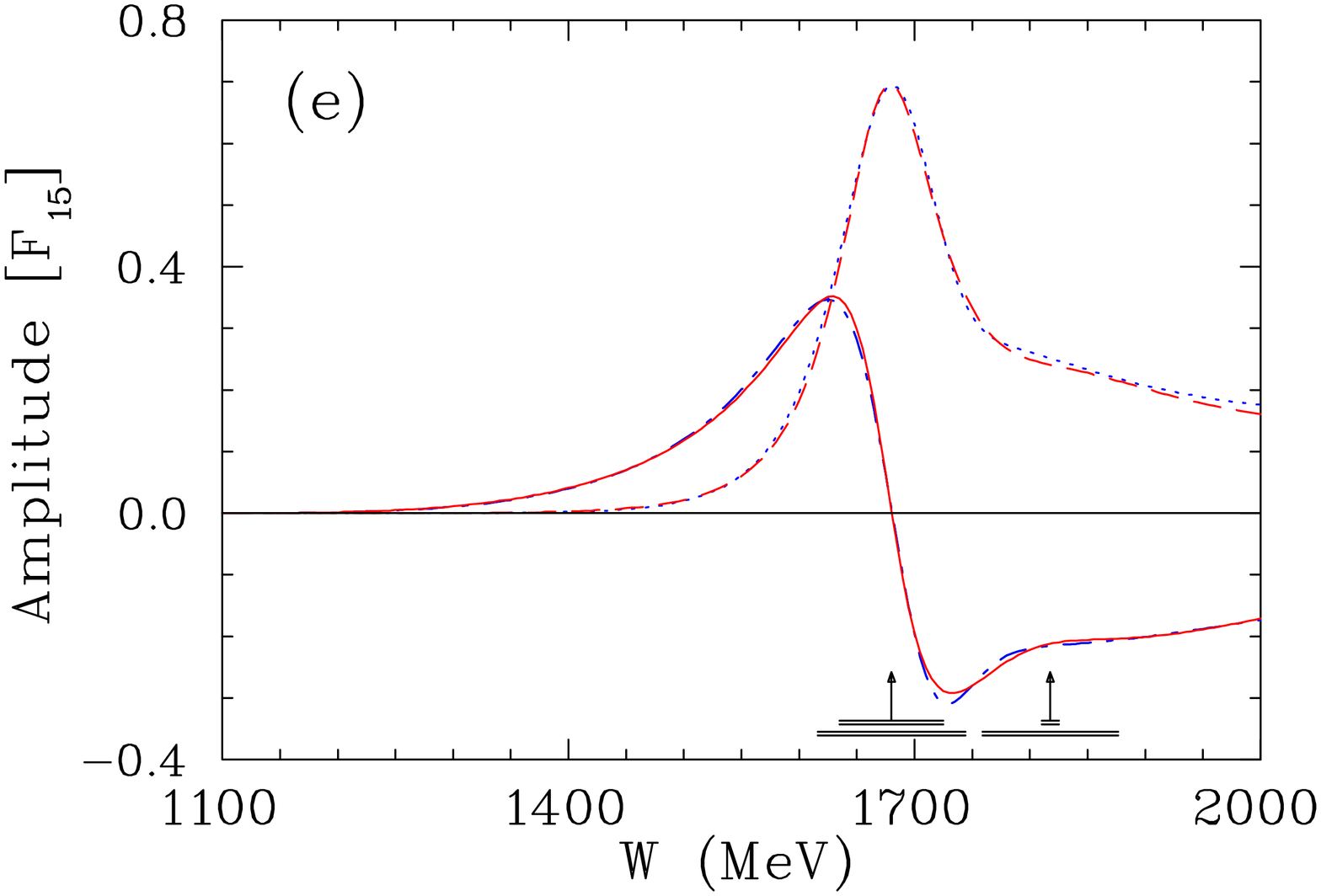}\hfill
\includegraphics[height=0.42\textwidth, angle=90]{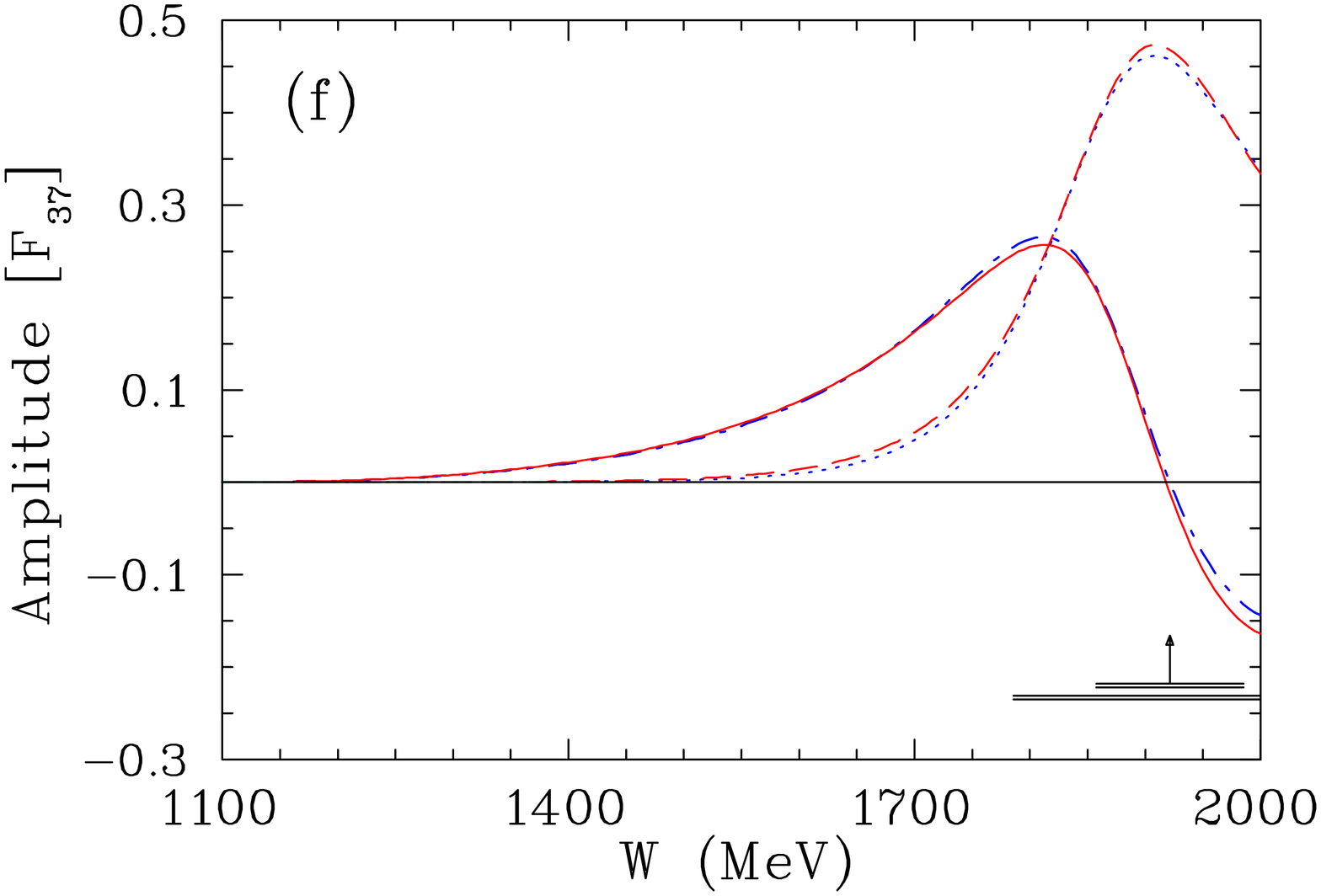}}
\caption{(Color online) Selected partial-wave amplitudes 
        (L$_{2I, 2J}$). Solid (dashed) curves give the
        real (imaginary) parts of amplitudes corresponding
        to the WI08~\protect\cite{SAID-website} solution.
        Dash-dotted (dotted) curves give the real (imaginary)
        parts of amplitudes corresponding to the XP08
        solution. (a) $S_{11}$, (b) $S_{31}$, (c) 
        $P_{11}$, (d) $D_{13}$, (e) $F_{15}$, and
        (f) $F_{37}$.  All amplitudes are dimensionless.
        Vertical arrows indicate resonance $W_R$ values and
        horizontal bars show full $\Gamma$ and partial widths
        for $\Gamma_{\pi N}$.} \label{fig:f1}
\end{figure*}
Given the $T$ matrix at some energy, $T(E)$, we can form the $S(E)$ 
matrix. We diagonalize this matrix using a standard routine to obtain
the eigenvalues $\{\lambda_i(E)\}_{i=1}^n$, where $n$ is the number 
of channels.

If the eigenvalues are nearly degenerate at some energy, it is
difficult to distinguish which eigenvalue corresponds to a given
eigenchannel, say $i$, since diagonalization of $S$ doesn't
preserve the eigenchannel ordering. The set of eigenvectors,
however, must be orthogonal at any energy; and, for continuous
partial wave amplitudes, the change of the eigenvector for a
given eigenchannel is small for nearby energies.

The eigenchannels are maintained using the following method. The
$S$ matrix is diagonalized at the initial energy, say
$E_1=1150$~MeV. We obtain $n$ eigenphases (where $n$ is the
number of channels included for the given partial wave),
$\lambda_1(E_1), \ldots, \lambda_n(E_1)$ and their corresponding
eigenvectors $v_1(E_1),\ldots,v_n(E_1)$. We wish to correlate the
eigenvalues and eigenvectors with a given eigenchannel throughout
the evaluation of the eigenvalues at higher energies, $E>E_1$.

Increasing the energy a small amount ($10-15$~MeV) to $E_2$, we 
again diagonalize the $S$ matrix and evaluate the 
$\lambda_1(E_2),\ldots,\lambda_n(E_2)$ and eigenvectors $v_1(E_2),
\ldots,v_n(E_2)$.

In order to track the eigenchannel, we evaluate the matrix of 
overlaps:
\begin{align}
	O_{ij}(E_1,E_2) &= v_i(E_1)^\dag v_j(E_2).
\end{align}
As $E_2\to E_1$, we have
\begin{align}
	\lim_{E_2\to E_1} O_{ij}(E_1,E_2) &= \delta_{ij},
\end{align}
which is just the statement that the eigenvectors are orthonormal.
For $E_2 - E_1 \simeq 10$~MeV, we identify the eigvenvalues according 
to the largest overlap in the set
\begin{align}
	\{ |O_{ij}(E_1,E_2)| \}_{j=1}^n.
\end{align}
Suppose, for example, that we have three channels and at the
energy $E_1$, we write the eigenvalues in the order:
\begin{align}
	\lambda_1,\lambda_2,\lambda_3.
\end{align}
And at energy $E_2$ for $i=1$, we find that
\begin{align}
	| O_{13}(E_1,E_2) | > | O_{11}(E_1,E_2) |
	> | O_{12}(E_1,E_2) | ,
\end{align}
then for energy $E_2$ we order the eigenvalue $\lambda_3$ first;
the ordering for the other eigenvalues is determined similarly.

\begin{figure*}[ht]
\centerline{
\includegraphics[height=0.42\textwidth, angle=90]{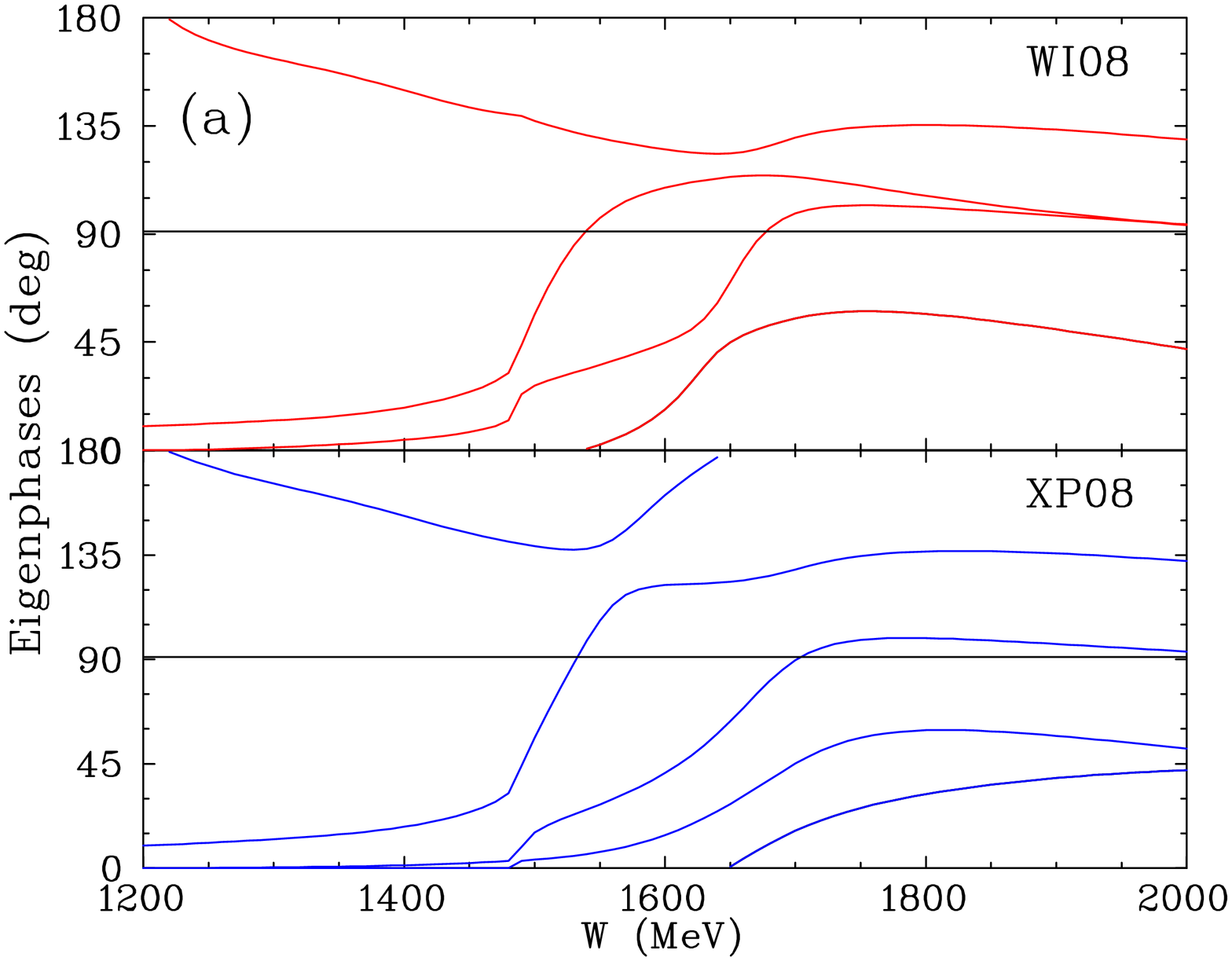}\hfill
\includegraphics[height=0.42\textwidth, angle=90]{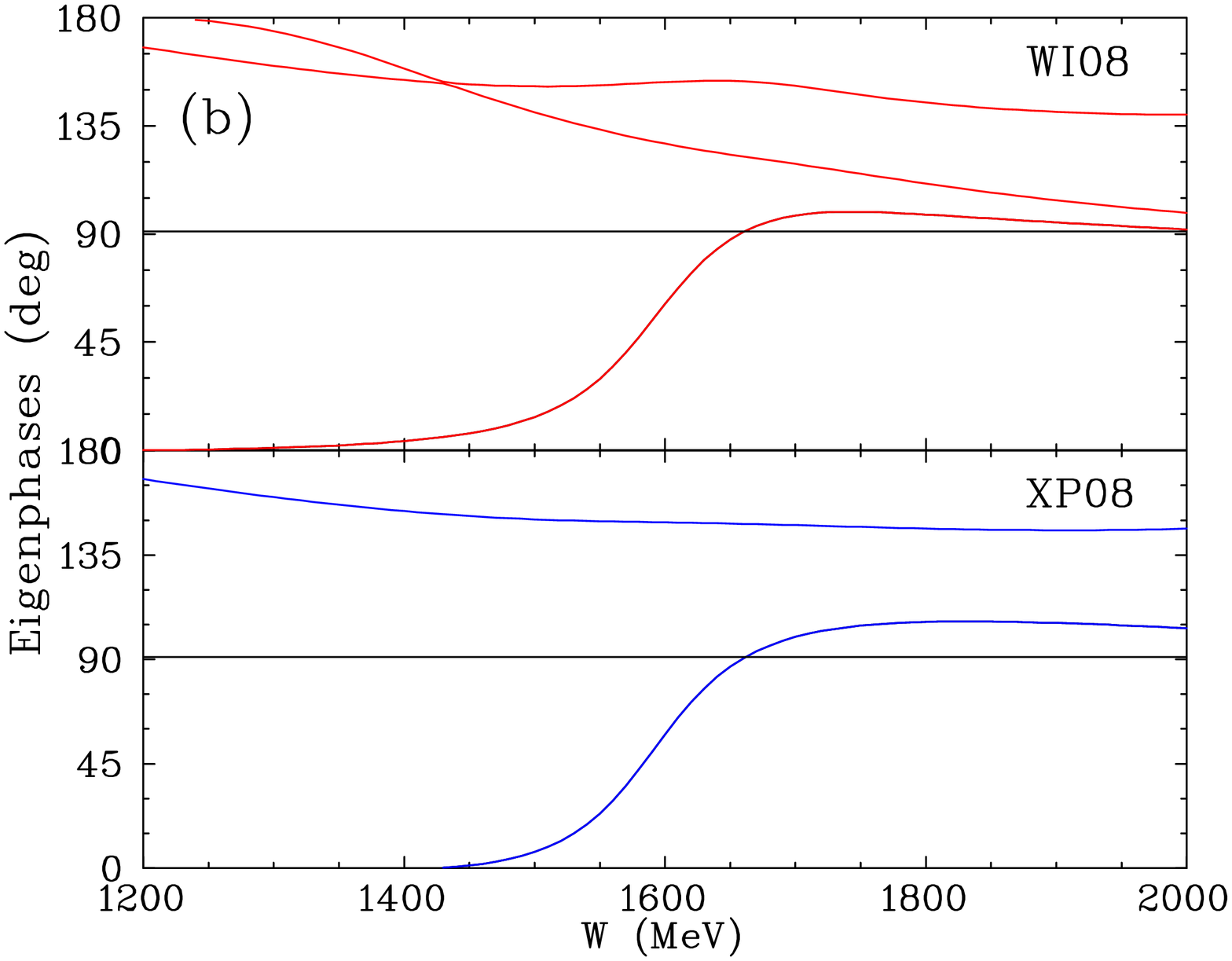}}
\centerline{
\includegraphics[height=0.42\textwidth, angle=90]{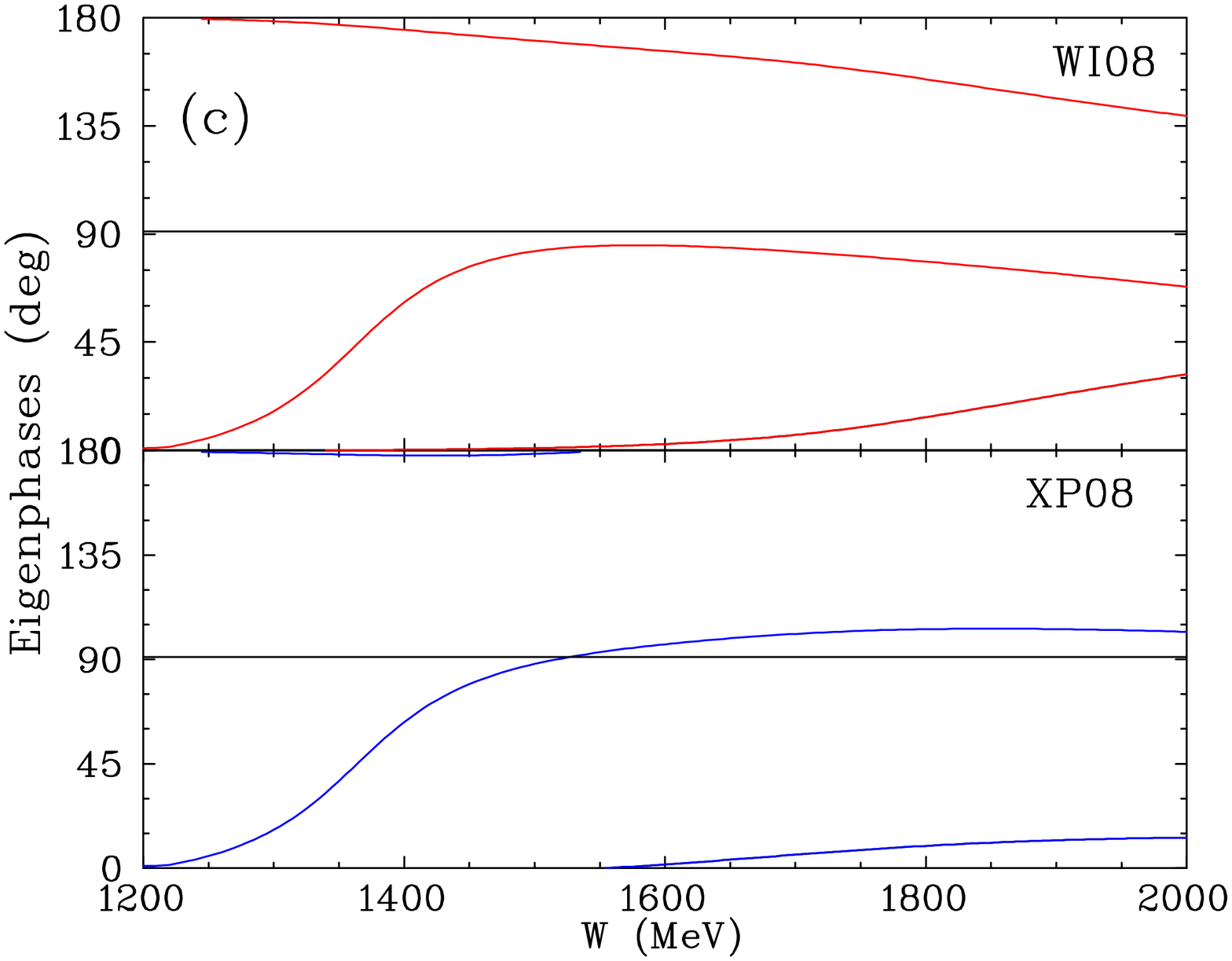}\hfill
\includegraphics[height=0.42\textwidth, angle=90]{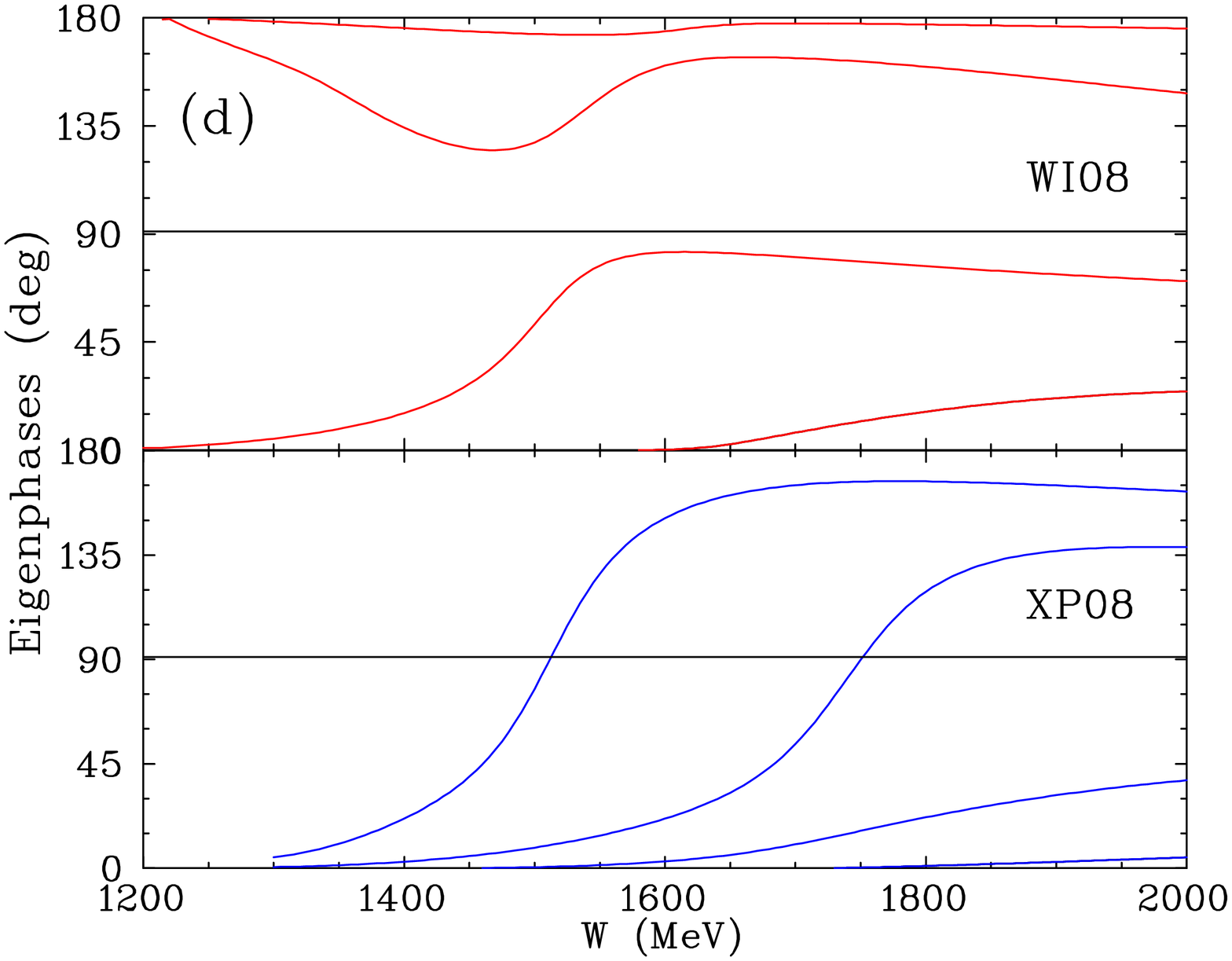}}
\centerline{
\includegraphics[height=0.42\textwidth, angle=90]{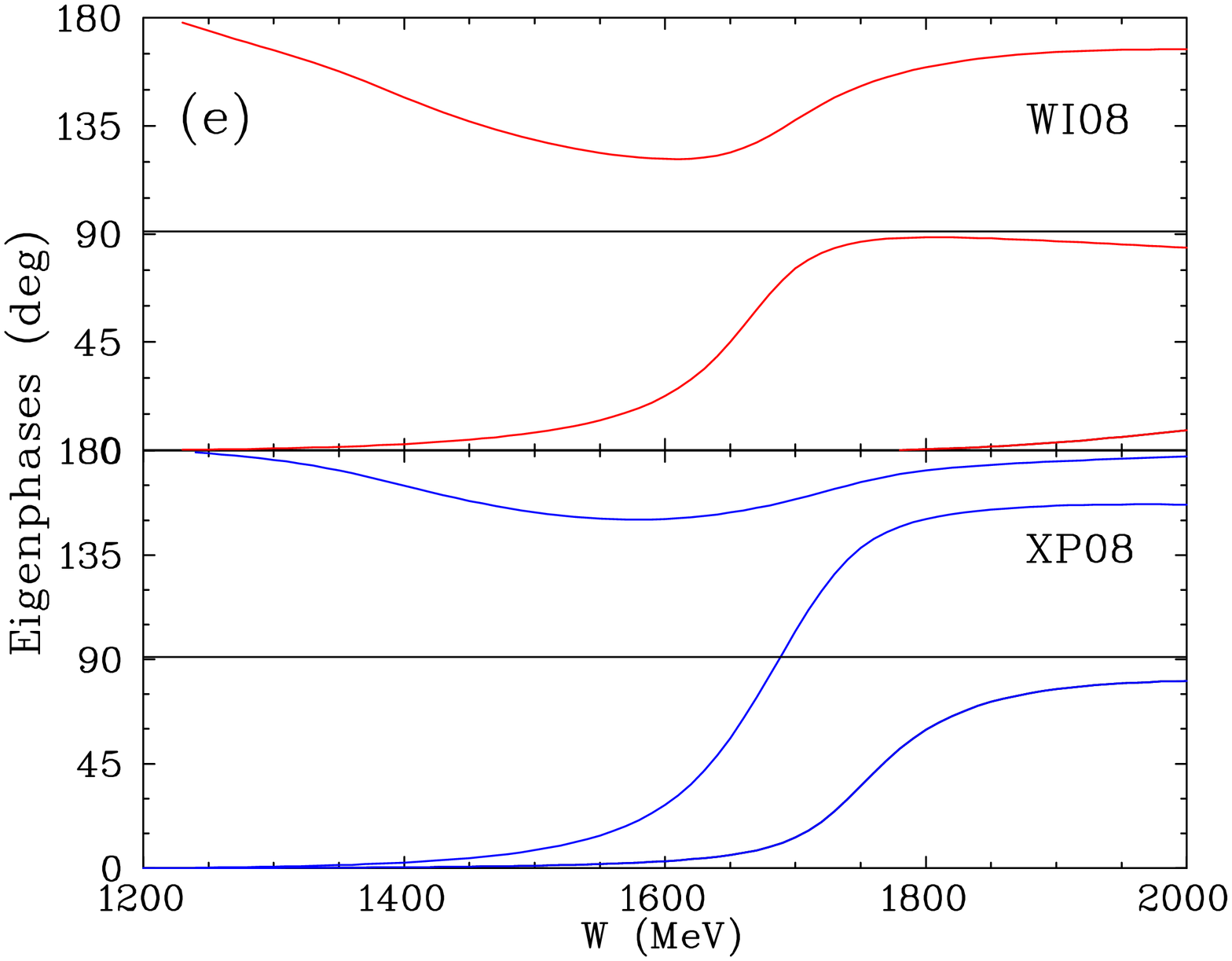}\hfill
\includegraphics[height=0.42\textwidth, angle=90]{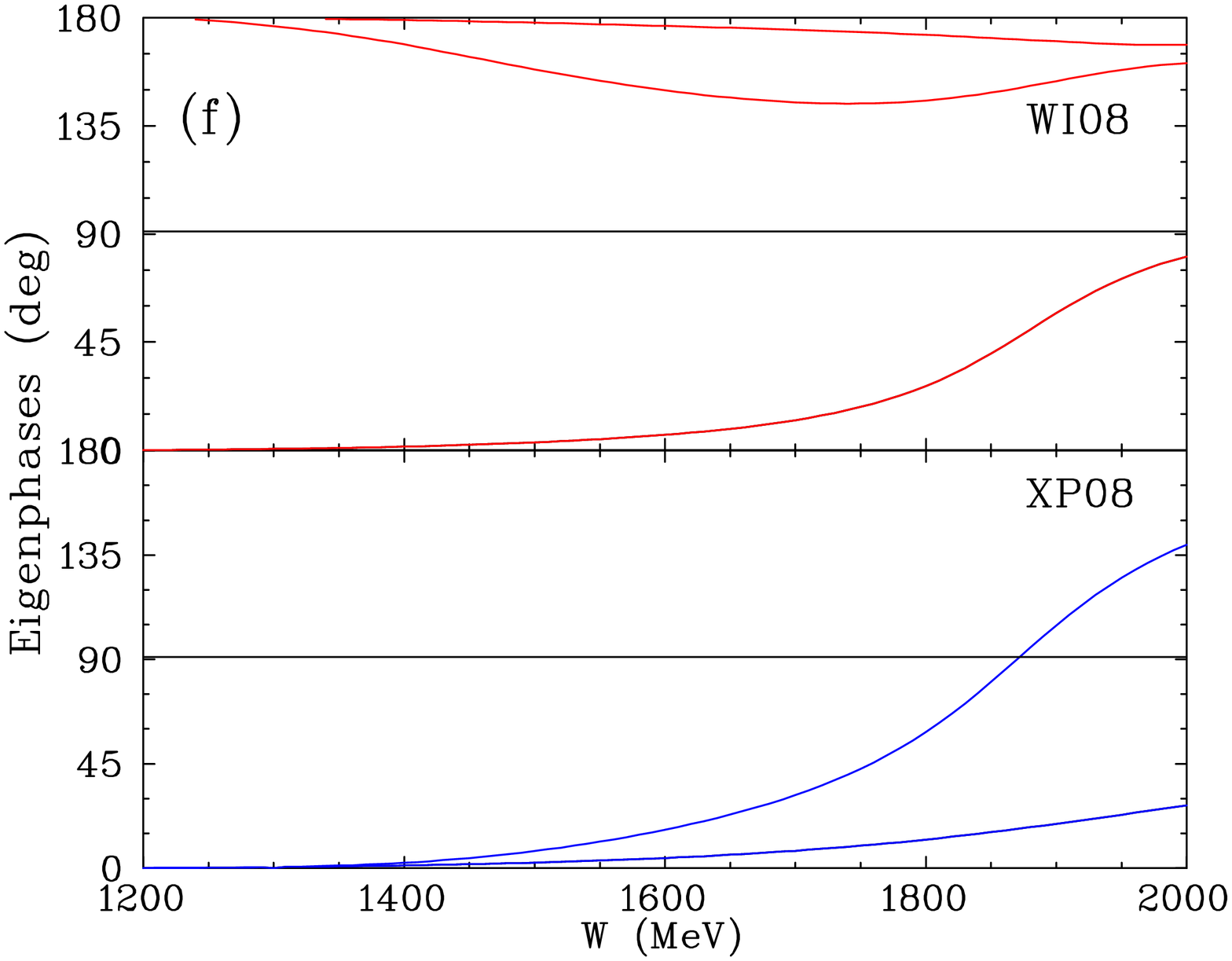}}
\caption{(Color online) Eigenphases.
	(a)  $S_{11}$, (b)  $S_{31}$, (c)  $P_{11}$, 
	(d)  $D_{13}$, (e)  $F_{15}$, and (f)  $F_{37}$.} 
	\label{fig:f2}
\end{figure*}
\begin{figure*}[ht]
\centerline{
\includegraphics[height=0.42\textwidth, angle=90]{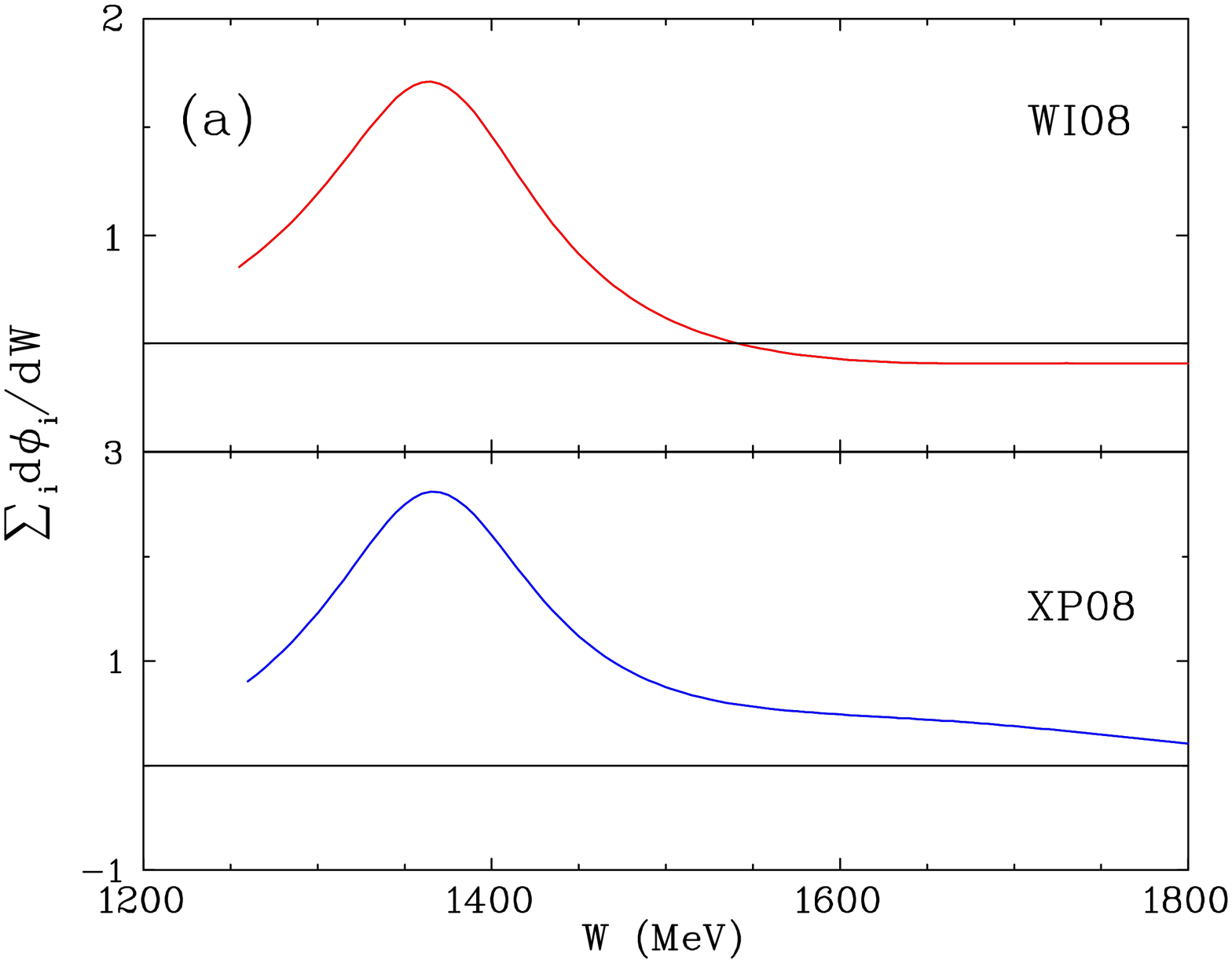}\hfill
\includegraphics[height=0.42\textwidth, angle=90]{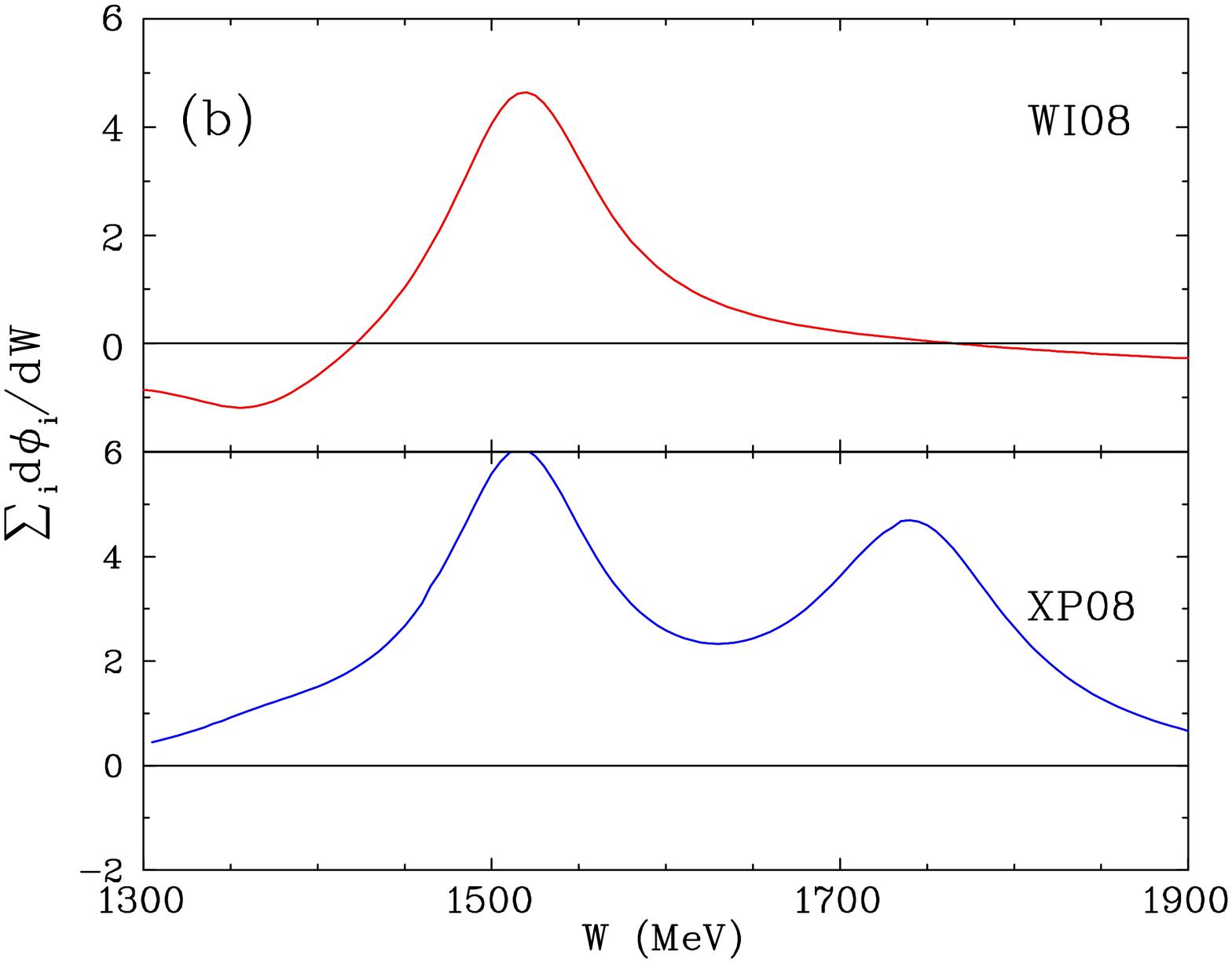}}
\caption{(Color online) Derivatives of eigenphases.
	(a) $P_{11}$ and (b) $D_{13}$.}
        \label{fig:f3}
\end{figure*}

\section{Results}
\label{sec:results}

\begin{table*}[ht]
\caption{Pole positions in complex energy plane of the $T$ matrix 
	for the $\pi N\to\pi N$ reaction. The functional forms 
	(see text) employed in the {\sc SAID} fits are compared 
	for selected partial waves. Each $T$ pole position is 
	expressed in terms of its real and imaginary parts 
	($M_R,-\Gamma_R/2$) in MeV. The second sheet pole is
        labeled by a $\dagger$. \label{tab:pole1}}
\begin{tabular*}
{\textwidth}{@{\extracolsep{\fill}}|c|ccc|ccc|}
\colrule
$\ell_{JT}$ & 
\multicolumn{3}{c}{{\sc WI08}} &
\multicolumn{3}{c}{{\sc XP08}} \\
\colrule
$S_{11}$ & $(1499,49)$  & $(1647,42)$ & $(1666,260)$
         & $(1538,65)$  & $(1675,58)$ & $(1690,121)$ \\
$S_{31}$ & $(1594,68)$  & &
         & $(1592,66)$  &  & \\
$P_{11}$ & $(1358,80)$  & $(1388,82)^\dagger$ & 
         & $(1358,80)$  & $(1387,80)^\dagger$ & $(1646,290)$ \\
$D_{13}$ & $(1515,55)$  &  &
         & $(1513,53)$  & $(1740,66)$ & $(1716,370)$ \\
$F_{15}$ & $(1674,57)$  & $(1779,138)$ & 
         & $(1672,70)$  & $(1734,61)$ & \\
$F_{37}$ & $(1883,115)$ & &
         & $(1874,119)$ & & \\
\colrule
\end{tabular*}
\end{table*}

The fits with (XP08) and without (WI08) explicit CM $K$-matrix
poles, in waves other than $(L_{2I,2J})$ $P_{33}$, are 
compared in Fig.~\ref{fig:f1}. Differences in the partial 
waves are slight, and the fit quality is comparable over the 
resonance region, each fit using a similar number of parameters. 
This feature of the $\pi N$ elastic scattering analysis seems 
quite stable.

In Fig.~\ref{fig:f2}, we have calculated the eigenphases 
corresponding to the full S-matrix. Only the $\pi N\to\pi N$ 
and $\pi N \to\eta N$ channels have been constrained by data.
The behavior of these phases does, however, provide an
interesting perspective on the emergence of resonance
structures in the fits. 

In the $S_{11}$ partial wave, both fits have two eigenphases
crossing 90 degrees, at 1535 and 1670~MeV for WI08, and at 
1530 and 1700~MeV for XP08. If one computes the usual Heitler 
$K$ matrix, as was done in Ref.~\cite{WAP}, $K$-matrix poles 
are found at these energies (since the unitary transformation, 
$U$ [Eq.\eqref{eqn:diagS}] diagonalizes $K$ simultaneously with 
$S$ and $K_{ii}=\tan\phi_i$). In the $S_{31}$ partial wave, a 
2- and 3-channel fit are compared, yielding identical crossing 
energies, again corresponding to a $K$-matrix pole (at 1655~MeV). 
Note that in the WI08 plot, two eigenphase nearly touch, but do 
not cross. 

In the $P_{11}$ plot, only one of the solutions has
a 90 degree crossing leading to a $K$-matrix pole. Note,
however, that the energy dependence of the eigenphase
crossing, and nearly crossing 90 degrees, is very
similar. This feature determines another measure of
resonance behavior, to be discussed below.

The $D_{13}$ eigenphases are quite different in the two
fits. In the WI08 fit, there are no 90 degree crossings,
while in XP08, we see two crossings. This hints at a 
different resonance structure, though the $\pi N$ 
$T$ matrices are nearly identical. 

In the $F_{15}$ and $F_{37}$ eigenphase plots, the
XP08 solution has a single crossing, whereas the WI08 
solution does not. Here also, a comparison of the 
eigenphases which cross, or come close to crossing, 90 
degrees have a similar energy dependence.

As has been noted previously\cite{Ceci:2008kt}, resonances may 
be associated with a single eigenphase crossing 90 degrees, 
and this will result in a $K$-matrix pole. However, a more 
robust measure (if a set of amplitudes is available) is given 
by the time-delay matrix~\cite{res}, which is proportional to 
the sum of energy derivatives of all eigenphases. Other factors, 
such as threshold openings can also produce rapid energy 
dependence.  Certainly the correct method of resonance 
identification requires the location
of poles in the complex energy plane on unphysical sheets close
to the physical region, which we demonstate below. Our employment
of the eigenphase approach illustrates the fact that resonance
structure may vary with the appearance of resonances in different
parametrizations without significantly altering the shape of the
$\pi N$ elastic amplitude. It is usually the case, however, that
such resonances are deep in the complex plane having large
widths.

In Fig.~\ref{fig:f3}, for illustration, we plot the sum of 
eigenphase energy derivatives for the $P_{11}$ and $D_{13}$. The 
peaks for $P_{11}$ are nearly identical and occur at about 1350~MeV, 
which (we will see) corresponds with the real part of the pole 
position. For the $D_{13}$, peaks corresponding the the 4-star 
state, near 1500~MeV, are closely aligned.  The second peak has 
almost no evidence in the $\pi N$ elastic amplitude. However, a 
large contribution to the (unfitted and therefore unconstrained) 
$\pi\Delta$ or $\rho N$ channels would result in the second peak. 

In Table~\ref{tab:pole1}, we compare the pole positions associated 
with resonance behavior in the plotted amplitudes. The third 
$S_{11}$ pole in XP08 closely resembles the structure found in
Ref.~\cite{sm95}, at $(1689,96)$~MeV, by scanning all partial
waves with an added Breit-Wigner contribution.  The very broad
$(1646,290)$ MeV $P_{11}$ state is similarly close to one found
in the SM90 fit~\cite{sm90}, at $(1636,272)$~MeV. Two extra poles 
were found in the $D_{13}$ partial wave for the XP08 solution 
compared to WI08. We do not intend to report the $(1716,370)$~MeV 
pole as a resonance but merely mention it here in connection with 
the present sensitivity study. Interestingly, the pole at
$(1740,66)$~MeV has its effect masked by a zero intervening
between the pole and real energy axis and therefore makes little
impact in the physical region.

\section{Conclusion}
\label{sec:concl}

We have reported a study of the parameterization dependence
of our $\pi N$ elastic amplitudes and resonance spectrum,
using a very different form for the CM $K$ matrix, with 
explicit poles in each partial wave. The partial-wave
amplitudes were found to be very stable with the change.

The eigenphase representation was introduced, mainly as
a novel approach to resonance identification, and because
it provides a more concrete example of properties discussed
in older works. This discussion also provides a 
continuation of the study started in Ref.~\cite{WAP}. 

The more formally correct extraction of pole positions has
revealed structures mainly found in earlier fits to the
$\pi N$ elastic scattering data. As the partial wave 
amplitudes have not changed significantly, the effects
of new resonances must be minimized through large widths,
intervening zeros, or small coupling to the $\pi N$ 
channel. 

In the SM90 fit, a study of the resonance spectrum was tried 
where, in addition to experimental data, the amplitudes from 
the KH~\cite{KH80} and CMB~\cite{CMB} analyses were added as 
soft constraints. A possible extension to the present work 
would be a re-examination of the resonance spectrum from a 
fit, with explicit CM $K$-matrix poles, constrained to more 
closely follow either the KH and CMB analyses, or a 
multi-channel analysis.  

\begin{acknowledgments}
This work was supported in part by the U.S. Department of Energy 
Grant DE-FG02-99ER41110. 
\end{acknowledgments}
\eject


\eject
\end{document}